\documentstyle[prd,aps,floats,epsfig]{revtex}
\tighten

\newcommand{\UNIT}[2]{{\mathrm{#1}^{#2}}}
\newcommand{\UUNIT}[2]{{\;\mathrm{#1}^{#2}}}

\newcommand{\GROUPE}[2]{{\mathrm{#1}({#2})}}

\newcommand{\ddd}{{\rm d}}
\newcommand{\ddpar}{{\rm \partial}}

\newcommand{\MAT}{{\rm m}}
\newcommand{\QUINT}{{\rm Q}}
\newcommand{\PL}{{\rm Pl}}
\newcommand{\REH}{{\rm reh}}
\newcommand{\CRIT}{{\rm c}}
\newcommand{\ADD}{{\rm add}}
\newcommand{\BROKEN}{{\rm bro}}
\newcommand{\BREAK}{{\rm SB}}
\newcommand{\OBS}{{\rm obs}}
\newcommand{\STRING}{{\rm S}}
\newcommand{\MIN}{{\rm min}}
\newcommand{\MAX}{{\rm max}}
\newcommand{\END}{{\rm end}}
\newcommand{\OSC}{{\rm osc}}
\newcommand{\SUGRA}{{\rm SUGRA}}
\newcommand{\RP}{{\rm RP}}

\newcommand{\DTRM}{{\rm D}}
\newcommand{\FTRM}{{\rm F}}

\newcommand{\HC}{{\rm h.c.}}  

\newcommand{\OO}{{ O}}

\newcommand{\ETAL}{{\it et al.}}

\begin{document}
\draft
\twocolumn[\hsize\textwidth\columnwidth\hsize\csname
@twocolumnfalse\endcsname

\title{Quintessence with two energy scales} 

\author{Philippe Brax}

\address{
  CERN, Theory division, Gen\`eve, Switzerland}

\author{J\'er\^ome Martin}

\address{ 
  Institut d'Astrophysique de Paris, 98 boulevard Arago, F--75014
  Paris, France}

\author{Alain Riazuelo}

\address{ 
  D\'epartement de Physique Th\'eorique, Universit\'e de Gen\`eve, 24
  Quai Ernest Ansermet CH--1211 Gen\`eve-4, Switzerland}

\maketitle

\begin{abstract} 
  We study quintessence models using low energy supergravity inspired
  from string theory. We consider effective supergravity with two
  scales $m_\STRING$, the string scale, and $m_\PL$, the Planck scale
  and show that quintessence naturally arises from a supersymmetry
  breaking hidden sector.  As long as supersymmetry is broken by the
  $F$-term of a Polonyi-like field coupled to the quintessence field
  in the K\"ahler potential we find that the Ratra-Peebles potential
  and its supergravity version are generic predictions. This requires
  that the string scale decouples from the Planck scale, $m_\STRING
  \ll m_\PL$. In the context of supergravity, the potential possesses
  a minimum induced by the supergravity corrections to the
  Ratra-Peebles potential at low redshifts. We study the physical
  consequences of the presence of this minimum.
\end{abstract}

\pacs{{\bf Preprint :} SPhT-T/01-044, CERN-TH/2001-112}

\vskip2pc]

\vskip1cm

\section{Introduction}

The cosmological constant problem is a long standing problem in
theoretical physics~\cite{Wein}. Although it first appeared in the
context of general relativity, it is now clear that this problem is
also deeply rooted in high energy physics. There exist several facts
which explain this link. The first is the observation that the
zero-point energy of quantum fields living in our universe effectively
acts as a bare cosmological constant of the Einstein equations. The
total contribution that we measure is therefore a combination of these
two terms. Since a naive argument leads to a zero-point energy
comparable to the Planck energy, and since astrophysical observations
tell us that the contribution is of the order of the critical energy
density, an extraordinary cancellation is needed if one wishes to
reconcile experiments with theories. A second fact became more evident
after the advent of supersymmetry (SUSY)~\cite{fer}. Indeed, in global
SUSY, the zero-point energy is guaranteed to vanish~\cite{BS}. This
raises the hope of finding a mechanism where the cosmological constant
would be exactly zero.  However, this explanation fails because SUSY
has to be broken in order to explain the heavy masses of the
superpartners. This means that the cosmological constant has to be at
least of the order of the SUSY breaking scale, i.e., at least of order
$1 \UUNIT{TeV}{}$, a value which still requires a very accurate
fine-tuning. On the other hand, when gravity is taken into account,
leading to supergravity (SUGRA), the fundamental state does not
necessarily have a zero energy. In that case, when SUSY is broken,
there is still the hope of finding a vanishing result. With the
present state of the art, this requires again a fine-tuning, and
cannot be derived from a fundamental principle. Therefore, the very
small (vanishing?) value of the cosmological constant remains a
mystery. The previous considerations led mant people to believe that
its value could be understood in the framework of the most promising
theory of high energy that was nowadays at disposal: string
theory~\cite{str}. It is widely believed that the only natural outcome
of string theory is that it must be zero~\cite{Witten}.  However, it
is clear that we are far from being in a situation where this
theoretical prejudice can be convincingly justified.
\par
An explanation for this puzzling question has recently become even
more necessary, since a combination of astrophysical observations,
including, among others, measurements of the Hubble diagram with type
Ia supernovae~\cite{SNIa}, seem to indicate that a form of dark energy
now dominates our universe; also see
Refs.~\cite{B98a,B98b,MAX1a,MAX1b,Teg,ZD,WCOS}. This has led to the
idea of quintessence. In this framework, the cosmological constant
vanishes exactly due to an as yet unknown mechanism, and a scalar
field is responsible for the acceleration of the expansion of the
universe~\cite{RP,Wett,FJ1,ZWS,SWZ}. In the same manner as for the
vanishing cosmological constant, it seems likely that the physical
nature of this field can be understood within the framework of string
theory, or at least within the framework of theories describing its
low energy limit. In this paper, we will adopt this point of view.
\par
Let us now recall some of the main ingredients of an effective SUGRA
description of string theory. The effective action describes the low
energy degrees of freedom which can be viewed as massless string
excitations. By computing string scattering amplitudes, one can build
an order by order perturbative expansion in the massless fields. This
perturbative expansion possesses two characteristic scales: the string
scale $m_\STRING$, and the compactification radius $R_{\rm c}$
springing from the necessary compactification from ten to four
dimensions. These two scales can be combined to form the Planck scale
$m_\PL$ which naturally appears at string tree level and parametrizes
the SUGRA expansion. The effective Lagrangian appears as a double
series in the string scale and the Planck scale.  In the context of
heterotic theory, both these scales are large and almost coincide.  In
new scenarios involving type I strings and $D$ branes there can be a
decoupling regime $m_\STRING \ll m_\PL$~\cite{ak}. We will show that,
in this context, the SUGRA potential introduced in
Refs.~\cite{bm_99,bm_99_2} is a prediction of the theory (for another
SUGRA model of quintessence, see Ref. \cite{CNR}).
\par
It is also of the utmost importance to understand which role SUSY
breaking plays in the determination of the shape of the quintessence
potential. As already mentioned above, particle physics experiments
require that SUSY must be broken at a scale at least of the order of
$1 \UUNIT{TeV}{}$. This is notoriously difficult to achieve in
explicit string models. Indeed, such a breaking must arise from
nonperturbative effects which are often difficult to control. Despite
the absence of convincing models of SUSY breaking, one can use a more
phenomenological approach and parametrize the SUSY breaking sector by
$F$ terms responsible for the breaking~\cite{bri}. This is the
approach we will follow. In particular, one usually assumes that SUSY
is broken by $F$ terms of the dilaton $S$ and the moduli fields $T^i$
measuring the compactification scales. In new type I models one can
also consider the blowup modes associated with the fixed points of
orbifolds~\cite{rigo}. This breaking is supposed to occur in a hidden
sector only gravitationally coupled to the visible sector. At the
SUGRA level one generally assumes that the cosmological constant
vanishes, requiring that $(S + \bar S)^2 \vert F_S \vert^2 + \sum_{i =
  1}^3 (T^i + \bar T^i)^2 \vert F_{T^i} \vert^2 = 3 m_{3 / 2}^2 /
\kappa$, where $m_{3 / 2}$ is the gravitino mass.  This relation is
imposed in order to cancel large contributions when the breaking scale
is of order of the a few $\UNIT{TeV}{}$, a mere 60 orders of magnitude
larger than the critical density of the universe.  In the context of
quintessence we will reconsider the previous relation and analyze
contributions which lead to quintessential potentials. In particular,
and in order to comply with the existence of an attractor for the
quintessence field, we will have to consider that quintessence arises
from a hidden sector. This guarantees that the very small mass of the
quintessence field does not lead to the existence of a long-range
fifth force~\cite{mas}. Moreover, we will see that quintessence can be
most easily achieved in a SUSY breaking sector.  In order to guarantee
that the sparticle masses do not depend strongly on the quintessence
field and therefore on the evolution of the universe, we find that
that the hidden sectors responsible for quintessence and the
superpartner masses must differ.
\par
This paper is organized as follows. In Sec.~\ref{sec_SUGRA}, we
quickly review how the effective SUGRA inspired from string theory can
be used to calculate the shape of the quintessence potential. The
details could have been dropped in a paper intended for high energy
physicists, but we think that they are useful in order to render this
paper self-consistent for a more general audience. In
Sec.~\ref{sec_RP}, we study the SUGRA model proposed in
Ref.~\cite{bm_99} with $m_\STRING \simeq m_\PL$ which leads to the
so-called Ratra-Peebles potential~\cite{RP} of Eq.~(\ref{potRP}). In
particular, we study the corrections to this model, and show that
observable quantities like the equation of state parameter and its
derivative are sensitive to these corrections. In Sec.~\ref{sec_gen},
we study the generic shape of a potential arising from effective SUGRA
where the assumption $m_\STRING \simeq m_\PL$ has been relaxed. We
prove that, at small redshifts, a generic form of the potential is
precisely the one found in Refs.~\cite{bm_99,bm_99_2}. In addition, it
is also established that the corrections no longer modify the shape of
the potential, which now really appears as a prediction and not as the
result of a particular model. In Sec.~\ref{sec_SUSY}, we study how
SUSY breaking by moduli fields can affect the form of the potential.
Again, it is found that the generic prediction is not changed.
Finally, since the SUGRA potentials generically possess a minimum, in
Sec.~\ref{sec_min} we study the observational consequences of this
fact. It is demonstrated that the quintessence field oscillates at the
bottom of its potential, but, depending on the precise depth of this
minimum, the field may or may not have begun its oscillations today.
It is also shown that in this framework it is unlikely that the
minimum of this potential can be put to zero.

\section{Model building and effective supergravity}
\label{sec_SUGRA}

One of the main advantages of the quintessence scenario is that the
coincidence problem can be solved, i.e., it is not necessary to
fine-tune the initial conditions at reheating in order to understand
why the dark energy starts dominating the matter content of the
Universe nowadays. This is due to the fact that for a potential with
the shape~\cite{RP}
\begin{equation}
\label{potRP}
V(Q) = \frac{M^{4 + \alpha}}{Q^\alpha}
\end{equation}
(typically), the Klein-Gordon equation possesses a (scaling)
solution~\cite{jpu} which is an attractor, also referred as a
``tracking solution''~\cite{ZWS}. This means that, whatever the
initial conditions are, in an allowed range encompassing more than
$100$ orders of magnitude, a given solution of the Klein-Gordon
equation will always tend toward the attractor before the present
epoch. When the field is on tracks, it satisfies the
equation~\cite{RP,ZWS}
\begin{equation}
\label{attra}
\frac{\ddd^2 V(Q)}{\ddd Q^2}
 = \frac{9}{2} H^2 \frac{\alpha + 1}{\alpha} (1 - \omega_\QUINT^2),
\label{atra}
\end{equation}
where $H$ is the Hubble parameter, and $\omega_\QUINT$ is the equation
of state parameter, i.e., the pressure to energy density ratio of the
scalar field. This is an important equation, because it allows us to
understand the different regimes undergone by the quintessence field
during cosmic evolution. Therefore, it can be used as a hint to which
kind of physics must be used in order to build a realistic and
successful model of quintessence. Equation~(\ref{attra}) has the
following consequences. First, it implies that the mass of the
quintessence field now is of the order of $H_0 \simeq 10^{- 43}
\UUNIT{GeV}{}$. Such a small mass entails that direct couplings
between the quintessence field and standard model fields, have to be
extremely suppressed. This suggests that the quintessence field
belongs to a hidden sector of the theory in order to avoid direct
couplings with the standard model fields, which would result in the
existence of a non-observed long-range interaction. Second, since the
second derivative of the potential is approximatively given by $\simeq
\rho_\QUINT / Q^2$ and since, when the field is about to dominate, we
also have $H^2 \simeq \rho_\QUINT / m_\PL^2$, we deduce that, at small
redshift, $Q \simeq m_\PL$. This means that supergravity effects will
be important at small redshifts, for example for the calculation of
the numerical value of the equation of state. In addition, we can also
estimate the value of the scale $M$ in the potential. One has
\begin{equation}
\label{M_val}
M \simeq (\rho_\CRIT m_\PL^\alpha)^{\frac{1}{4 + \alpha}}, 
\end{equation}
where $\rho_\CRIT$ is the critical energy density. Third, we know
that, initially, the value of the energy density of the dark energy
must be between the value of the background energy density at
reheating, i.e., $\rho_\REH \simeq 10^{61} \UUNIT{GeV}{4}$, and the
background energy density today, i.e. $\rho_\CRIT \simeq 10^{- 47}
\UUNIT{GeV}{4}$.  Starting from this range guarantees that the field
$Q$ will join the attractor before now. This range for the initial
energy density of the dark energy corresponds to very small values of
the field itself, $Q \ll m_\PL$.  More precisely, if the field starts
at rest, we initially have $10^{- 108 / \alpha } m_\PL \le Q \le
m_\PL$.  Unless the field starts with an energy density of same order
as today's critical density, this implies that supergravity effects
are negligible at the beginning of the evolution, and that this epoch
can be well described by means of a globally supersymmetric theory.
\par
Having identified the orders of magnitude of the value of the scalar
field throughout the cosmic evolution, we can study the physics which
is necessary to describe these different regimes. We are going to
consider an effective SUGRA theory and the constraints imposed by
quintessence. In particular, as mentioned above, we assume explicitly
that the quintessence field belongs to a hidden sector of the theory.
We assume that the effective action is a double series expansion in
the Planck mass and in the string scale. The Planck mass $m_\PL$ and
the string scale $m_\STRING$ are {\it a priori} two independent
scales. The only experimental constraint is that $m_\STRING > 1
\UUNIT{TeV}{}$ in order not to be in conflict with the measurements
performed by the accelerators. In heterotic string inspired models it
was often assumed that $m_\STRING \simeq m_\PL$, because of the
constraints on the perturbative unification scale.  However, recently,
models where the string scale is much lower than the Planck scale were
proposed~\cite{ak}.  Generically, these two scales are linked by the
equation
\begin{equation}
m_\PL^2 = m_\STRING^8 V_6,
\end{equation}
where $V_6$ is the volume of the six compactified dimensions.  The
constraint that $m_\STRING> 1 \UUNIT{TeV}{}$ translates into a
constraint on the volume $V_6 < 10^{14} \UUNIT{GeV}{- 6}$. As
mentioned above, it was recently shown that some of the compactified
dimensions can be large (in comparison to the Planck length) resulting
in a string scale much lower than the Planck scale. In this paper, for
the moment, we leave $m_\PL$ and $m_\STRING$ free. We will discuss the
different cases later on.
\par
Because of the large value $Q \simeq m_\PL$ of the quintessence field
today, it appears that one would need a full understanding of the
complete SUGRA action, i.e., one would need to take into account all
the $Q / m_\PL$ and $Q / m_\STRING$ terms in the Lagrangian. As a
result, one would expect that an appropriate description of
quintessence requires an understanding of nonperturbative effects
either at the field theoretical level or even at the string level. In
the following we shall use a more modest approach and remain within a
perturbative setting where the Lagrangian is expanded in inverse mass
powers. We will pay particular attention to the sensitivity of the
physical observables to the degree of the truncated perturbative
series.  In particular we will comment on the stability of the
physical observables under a change of the truncation degree. The only
nonperturbative inputs will be the SUSY breaking parameters.
\par
Let us first consider the early universe evolution of the quintessence
field. Setting the initial conditions at reheating, after inflation,
implies that for most of the time the quintessence field takes values
which are negligible with respect to the Planck mass. We assume that
the expectation values of the other fields are also negligible in
comparison with the Planck mass. This means that, in this context, the
most general Lagrangian is given by the $N = 1$ (global) SUSY
Lagrangian [i.e., the $N = 1$ SUGRA Lagrangian where terms of order
$\OO(m_\PL^{- 1})$ are neglected]
\begin{eqnarray}
\label{SUSYL}
{\cal L}
 & = &   \int \ddd^4 
         \theta K (\Phi^{i k  \dagger} e^{2 g_k V_k}, \Phi ^{i k
}) 
       + \int \ddd^2 
         \theta [W (\Phi^{i k}) + \HC] 
\nonumber \\ & &
       + \sum_k \int \ddd^2 
         \theta \sum_{a b} f_{a b} (\Phi^{i \ell })
         [W_{k a}{}^{\alpha} W_{k b \alpha} + \HC] 
\nonumber \\ & &
       + \sum_{k \in \GROUPE{U}{1}} \xi_k \int \ddd^4 \theta V_k .
\end{eqnarray} 
Because of the possible large hierarchy between the string scale and
the Planck scale, the K\"ahler potential $K$ and the superpotential
$W$ are now series in the inverse string scale. Let us now focus on
the hidden sector containing the quintessence field.  For simplicity,
and since it does not change our general argument, we will take $f_{a
  b} (\Phi^{i \ell }) = \delta_{a b}$. Let us describe this Lagrangian
in more detail. In the previous expression, $\Phi^{i k} (x^\kappa,
\theta, \bar{\theta})$ is a chiral superfield and $V_k (x^\kappa,
\theta, \bar{\theta})$ is a vector superfield which can be written in
terms of components as
\begin{eqnarray}
\Phi^{i k} (x^\kappa, \theta ,\bar{\theta})
 & = &   \varphi^{i k} (x^\kappa)
       + \sqrt{2} \theta \psi^{i k} (x^\kappa)
       + \theta^2 F^{i k} (x^\kappa) , \\ 
V_k (x^\kappa, \theta , \bar{\theta})
 & = & \sum_a \biggl[ - \theta \sigma^\mu 
                       \bar{\theta} V_{k a \mu} (x^\kappa)
\nonumber \\ & &
                     + i \theta \theta 
                         \bar{\theta} \bar{\lambda}_{k a} (x^\kappa)
                     - i \bar{\theta} \bar{\theta} 
                         \theta \lambda_{k a} (x^\kappa)
\nonumber \\ & & 
                     + \frac{1}{2} \theta \theta 
                       \bar{\theta} \bar{\theta} D_{k a} (x^\kappa)
              \biggr] T_{a k} 
   \equiv \sum_a V_{k a} T_{a k} ,
\end{eqnarray}
where the vector superfield has been written in the Wess-Zumino gauge.
We assume that the above Lagrangian is invariant under the gauge group
$G$ acting on the $k$ indices of the chiral superfields:
\begin{equation}
\label{group}
G = \prod_k G_k \times \GROUPE{U}{1}_X ,
\end{equation}
This gauge group might become strongly coupled and lead to SUSY
breaking via gaugino condensation. In the previous expressions, $k$ is
a group index, i.e., $V_k$ is the superfield charged under the group
$G_k$. Under this group $G_k$, many chiral superfields can be charged.
The index $i$ in $\Phi^{i k}$ labels the different superfields that
are charged under the group labeled by the index $k$. The matrices
$T_{a k}$ are the generator of the gauge group $G_k$ and the index $a$
runs from $1$ to $\dim(G_k)$.  In the third term of the above
Lagrangian, $W_{k a \alpha}$ is given by $W_{k a \alpha} \equiv - (1 /
4) \bar{D} \bar{D} e^{-V_{k a}} D_{\alpha} e^{V_{k a}}$, where $D$ is
the supersymmetric derivative. The extra $\GROUPE{U}{1}_X$ is an
anomalous Abelian factor associated with a Fayet-Iliopoulos term in
the Lagrangian in order to cancel the anomaly by the Green-Schwarz
mechanism~\cite{gs}.  In heterotic string theory there is a single
anomalous $\GROUPE{U}{1}_X$~\cite{dw}. In type I string theories there
may be several anomalous $\GROUPE{U}{1}$'s depending on the geometry
of the compactifying manifold~\cite{rigo}. The last term in the
Lagrangian represents the Fayet-Iliopoulos term where $\xi_k$ is a
constant different for each group, provided that this group is
$\GROUPE{U}{1}$.  {\it A priori}, the scale given by the
Fayet-Iliopoulos term is expected to be of the order of the string
scale. This is the case in the heterotic string theory for the unique
Fayet-Iliopoulos term for the anomalous $\GROUPE{U}{1}_X$.  In type I
string theories the Fayet-Iliopoulos terms are associated with the
blowing up moduli of orbifold singularities in the compactification
space. Their values parametrize a flat direction with no potential,
and are therefore left unfixed at the perturbative level of string
theory.
\par
Once the K\"ahler function $K$ and the superpotential $W$ of the
hidden sector are given, the Lagrangian [Eq.~(\ref{SUSYL})] is
completely fixed. In particular, the scalar potential can be
calculated. It contains two contributions: one comes from the $F$
terms and the other comes from the $D$ terms. Explicitly, the
potential is given by
\begin{equation}
\label{scalpot}
V = V_\FTRM + V_\DTRM
  =   K_{\bar{A} B} F^{\bar{A} \dagger} F^B
    + \frac{1}{2} \sum_{k a} D_{k a} D_{k a}  
\end{equation}
in the low energy limit.  In the previous expression, we introduced a
collective index $A \equiv (i k)$. The metric $K_{\bar{A} B}$ and the
field $F_A$ can be expressed as
\begin{equation}
\label{Fcara}
K_{\bar{A} B} = \frac{\ddpar^2K}{\ddpar\varphi^{A \dagger}
\ddpar\varphi^B}
,\quad 
F^A = - K^{\bar{B} A} 
        \frac{\ddpar W^\dagger}{\ddpar\varphi^{\bar{B} \dagger}},
\end{equation}
respectively, and the $D$ term is given by
\begin{equation}
\label{Dcara}
D_{k a} = - \frac{\xi_k}{2} 
          - \sum_i g_{k} \frac{\ddpar K}{\ddpar\varphi^{i k \dagger}}
                   T_{k a} \varphi^{i k \dagger}.
\end{equation}
In the last equation, we have assumed that the gauge group considered
is $\GROUPE{U}{1}$ otherwise the expression would be the same except
that the Fayet-Iliopoulos term would not be present.
\par
We now assume that SUSY is not broken by the $D$ terms. This implies
that $\left< D_{k a} \right> = 0$. If $k$ corresponds to a group which
is $\GROUPE{U}{1}$, this means that one (or many) of the scalar fields
acquire a nonvanishing vacuum expectation value, according to
\begin{equation}
\label{vev}
\xi_k
 = - 2 \sum_i g_k \left<\frac{\ddpar K}{\ddpar\varphi^{i k \dagger}}
                                 T_{k a} \varphi^{i k \dagger}
                  \right>
,\quad 
k \in \GROUPE{U}{1} .
\end{equation}
The generators $T_{k a}$ give the charges of the fields under the
considered $\GROUPE{U}{1}$.  Typically, one expects $\left< \varphi^{i
    k} \right> \simeq \sqrt{\xi_k}$. This means that the
$\GROUPE{U}{1}$ gauge symmetries are broken at that scale.  In the
heterotic case this fixes the breaking at the GUT scale while in the
type I models the breaking scale is not specified as it is a modulus.
We conclude that the $D$ part of the scalar potential vanishes, i.e.,
\begin{equation}
V_\DTRM = 0.
\end{equation}
The nonzero contributions to the potential come from the $F$ terms.
\par
The previous considerations are valid at very high redshift.  However,
at small redshift, one needs to take into account the effects of
SUGRA, since the values of the quintessence field are not negligible
compared to the Planck mass. In SUGRA, the form of the scalar
potential is modified, and reads
\begin{equation}
\label{sugraV}
V = \frac{1}{\kappa^2} e^G (G^A G_A - 3) + V_\DTRM,
\end{equation}
where $\kappa =8\pi /m_\PL^2$ and $G \equiv \kappa K + \ln(\kappa^3
|W|^2)$. In the previous expression, $G_A$ is defined by
\begin{eqnarray}
\label{defG}
G_A
 & \equiv &   \frac{\ddpar G}{\ddpar \varphi^A} 
   =          \kappa \frac{\ddpar K}{\ddpar \varphi^A} 
            + \frac{1}{W} \frac{\ddpar W}{\ddpar \varphi^A},
\\
G_{\bar{A}}
 & \equiv &   \frac{\ddpar G}{\ddpar \varphi^{\bar{A} \dagger}} 
   =          \kappa \frac{\ddpar K}{\ddpar \varphi^{\bar{A} \dagger}} 
            + \frac{1}{W^\dagger}
              \frac{\ddpar W^\dagger}{\ddpar \varphi^{\bar{A} \dagger}},
\end{eqnarray}
and the indices are raised and lowered with the help of the following
metric:
\begin{equation}
G_{\bar{A} B}
 \equiv \frac{\ddpar^2 G}
             {\ddpar \varphi^{\bar{A} \dagger} \ddpar \varphi^B}
 =      \kappa K_{\bar{A} B}.
\end{equation}
The other terms in $G$ cancel out because the superpotential is a
holomorphic function. {\it A priori}, this potential is no longer
positive definite. In particular, there is a negative contribution
coming from the superpotential.
\par
Let us come to grips with the quintessence potential more precisely.
According to the previous discussion, we only focus our attention to
the $F$ part of the scalar potential. A first attempt to derive the
Ratra-Peebles potential [Eq.~(\ref{potRP})] from first principles was
made in Ref.~\cite{Bine} and then in Refs.~\cite{bm_99,bm_99_2}. In
order to see clearly the difference between this approach and the
approach advocated in the present paper, we first quickly review the
results obtained in Refs.~\cite{bm_99,bm_99_2}. Then we will study in
detail new properties of the model presented in Ref.~\cite{bm_99_2}.
We will argue that these new properties are in fact a problem, and we
will see how, generically, they can be avoided in the context of
theories with two different scales.

\section{A SUGRA model leading to the Ratra-Peebles potential}
\label{sec_RP}

In the model presented in Refs.~\cite{bm_99,bm_99_2}, it is assumed
that $m_\STRING \simeq m_\PL$. Contrary to the strategy used in the
Sec.~\ref{sec_gen}, which is to see which kind of potential is
obtained from a generic theory, the idea utilized in
Refs.~\cite{bm_99,bm_99_2} was to study the required properties of the
theory such that the desired potential (typically the Ratra-Peebles
potential) is the result of the calculations described above. Below,
we improve the presentation of the model of
Refs.~\cite{bm_99,bm_99_2}, in particular we describe it in a language
closer to high energy physics than the one used in
Refs.~\cite{bm_99,bm_99_2}.
\par
We assume that there are three sectors in the theory. One of them is
the observable sector where all the known particles and their
superpartners live and the two other sectors are hidden. The first
hidden sector is the ``quintessence sector'' already mentioned above,
where the quintessence field lives. The second one is the ``broken
sector'', introduced such that SUSY should be broken in a satisfactory
manner. We have seen previously that, generically, due to the presence
of a Fayet-Iliopoulos term and to the vanishing of the potential
coming form the $D$ terms, at least one scalar field acquires a
nonvanishing vacuum expectation value. Let us call this field $Z$.
Thus we have $\left< Z \right> \neq 0$. This field belongs to the
quintessence sector. In addition, this sector is required to contain a
field $Y$ such that $\left< \partial_YW\right> \neq 0$. This field is
similar to a Polonyi field~\cite{Po}, although we do not assume that
the superpotential is linear in this field. We also assume that
$\partial_YW$, i.e., $F_Y$ in global SUSY, does not depend on $Q$. The
K\"ahler potential and the superpotential of
Refs.~\cite{bm_99,bm_99_2} have the forms
\begin{eqnarray}
\label{KWplb}
K &=& \frac{1}{m_\PL^{2 p}} \vert Y \vert^2 (Q\bar{Q})^p + 
\hat{K} (\vert Y \vert ^4 , \dots , \Phi_\QUINT, \Phi_\BROKEN, 
\Phi_\OBS), \\
\label{KWplb2}
W &=& Y Z^2 + \hat{W}_\QUINT(\Phi_\QUINT)+W_\BROKEN(\Phi_\BROKEN)
+W_\OBS (\Phi_\OBS),
\end{eqnarray}
where $\Phi_\QUINT$, $\Phi_\BROKEN$, and $\Phi_\OBS$ denote
superfields in the quintessence, broken and observable sectors
respectively.  $W_\BROKEN$ and $W_\OBS$ are the superpotentials in the
broken and observables sectors. $W_\QUINT \equiv Y
Z^2+\hat{W}(\Phi_\QUINT)$ is the superpotential in the quintessence
sector. We have $\langle W_\QUINT\rangle =\langle W_\OBS \rangle =0$
but $\langle W_\BROKEN \rangle \neq 0$. The condition $\langle
W_\QUINT\rangle =0$ guarantees that the SUGRA quintessence potential
is positive definite. Then, in the context of global SUSY, the scalar
potential is $V(Q) = m_\PL^{2 p} \vert F_Y \vert^2 / Q^{2 p}$, i.e.,
the Ratra-Peebles potential . We see that a crucial point in the
argument is the vanishing of the term $\vert Y \vert^2$ in the series
defining the K\"ahler potential.  Although this concerns only one term
in the complete series, this should probably be considered as an
unwanted fine-tuning, since there is no fundamental reason to expect
that this term must be absent in a generic theory. In addition, since
$M^{4 + \alpha} \simeq \rho_\CRIT m_\PL^\alpha$, one has
\begin{equation}
\label{breakmod}
\vert F_Y \vert^2 =\langle Z^2\rangle ^2 \simeq \rho_\CRIT,
\end{equation}
which fixes the scale at which SUSY is broken in the quintessence
sector.  We see that this scale is very small in comparison with the
``natural scale'' of SUSY breaking, i.e., $\simeq 1 \UUNIT{TeV}{}$.
\par
Actually, this is the main motivation for introducing two hidden
sectors. It is convenient to break SUSY in a hidden sector since, from
a phenomenological point of view, it seems difficult to break SUSY in
the observable sector. Indeed, for example, a spontaneous breaking
mechanism in the observable sector like the O'Raifeartaigh
mechanism~\cite{Ra} would not lead to a spectrum in accordance with
the constraints on the masses of the superpartners. Conversely, if the
hidden sector contains a Polonyi field $P$ (not the same as the one
contained in the quintessence sector, see above) such that $\langle
F_P \rangle = m_\BREAK^2$ and if the cosmological constant problem is
assumed to be solved (as it is always the case when one discusses
quintessence, see the introduction) then $m_\BREAK^2\simeq
m_{3/2}m_\PL$, where $m_{3/2}$ is the gravitino mass. This will give a
mass of order $m_{3/2}$ to the superpartners. Since we expect
$m_{3/2}\simeq 1 \UUNIT{TeV}{}$, this implies $m_\BREAK \simeq 10^{10}
\UUNIT{GeV}{}$ and $ \langle F_P\rangle \simeq 10^{20}
\UUNIT{GeV}{2}$, a value far from $F_Y$. Therefore, it is necessary
that the observable sector should be different from the broken sector
in order to have a correct spectrum, and it is also necessary that the
quintessence sector should be different from the broken sector in
order to have a value for $m_\BREAK$ of the correct order of
magnitude. In addition, the quintessence sector cannot be the
observable sector, since this would imply the presence of a long range
fifth force not seen in the data. In order to obtain the potential
which is valid not only at the beginning of the evolution but
everywhere, we need to insert the K\"ahler potential and
superpotential given in Eqs.~(\ref{KWplb}) and~(\ref{KWplb2}) in the
equation giving the scalar potential in SUGRA [Eq.~(\ref{sugraV})]. We
find that the only contributions which lead to nonvanishing terms in
the scalar potential are
\begin{equation}
\label{GGY}
G_{\bar{Y} Y} = \kappa K_{\bar{Y} Y}
,\quad 
F_Y\equiv -\frac{\ddpar W}{\ddpar Y} -\kappa WK_Y \neq 0,
\end{equation}
where $W$ stands for the total superpotential\footnote{Throughout the
  paper the auxiliary $F$ fields are given by ${\cal F}=e^{\kappa K/2}
  F$, where $F$ is defined by the second equation of
  Eqs.~(\ref{GGY}).}.  The vacuum expectation value of the last term
is in fact just $F_Y=-\partial W/\partial Y$. This is due to the
vanishing of the Polonyi-like field $\langle Y \rangle =0$ and the
quadratic dependence of the K\"ahler potential on $Y$. Finally, we
arrive at a positive definite expression
\begin{equation}
V = e^{\kappa K} K^{Y \bar Y}\langle Z^2\rangle ^2,
\end{equation}
where we have used the fact that the $D$ terms are not modified in
SUGRA and that, as a consequence, $\left< V_\DTRM \right> = 0$.  The
main difference comes from the exponential factor which represents the
SUGRA corrections. However, we do not have yet reached our main goal
because the kinetic term of $Q$ is still nonstandard.  Indeed, since
we are now in a regime where $Q \simeq m_\PL$, we can no longer
neglect the higher order terms in Eqs.~(\ref{KWplb})
and~(\ref{KWplb2}) and thus $K_{Q \bar Q} \neq 1$. The K\"ahler
potential evaluated at the minimum of the potential for the matter
fields reads
\begin{eqnarray}
\label{KQ}
K (Q, \left<Y\right> , \left<Z\right>, \left<\varphi^{i k}\right>)
 & = & \hat{K} (Q, \left< Z \right>, \left< \varphi^{i k} \right>) 
\nonumber \\
 & = & \sum_{n = 1}^\infty \frac{c_{2 n}}{m_\PL^{2 (n - 1)}}Q^{2 n} ,
\end{eqnarray}
where we have fixed the other hidden sector fields to their vev. This
means that the coefficients $c_{2n}$ are functions of $\langle Z
\rangle $. In a regime where $Q \ll m_\PL$, only the first term will
be important, and leads to a canonical kinetic term for quintessence
(with $c_2 = 1$). Therefore, the potential obtained in the context of
global SUSY is not modified by a redefinition of the field.  Closer to
the Planck scale the contributions from the other terms become
non-negligible.  To deal with this problem, we define a new scalar
field $\tilde{Q}$ such that
\begin{equation}
\label{staQ}
\ddd \tilde{Q}
 = \sqrt{2 K_{Q \bar Q}} \ddd Q 
 \Rightarrow 
\tilde{Q}
 = \int \ddd Q \sqrt{2 {K_{Q \bar Q}}}
 \equiv f(Q),
\end{equation}
where the function $f(Q)$ has been obtained by quadrature. The field
$\tilde{Q}$ has a standard kinetic term. Expressing $Q = f^{-1}
(\tilde{Q})$, we obtain the SUGRA potential
\begin{equation}
\label{sugraVQ}
V(\tilde{Q})
 = e^{\kappa K[f^{- 1} (\tilde{Q})]} 
   \frac{\langle Z^2\rangle ^2}{[f^{-1} (\tilde{Q})]^{2 p}} .
\end{equation}
{\it A priori}, any function $f(Q)$ is allowed. When $Q \ll m_\PL$,
the form of the function $f$ is irrelevant, since we know from the
previous SUSY considerations that the potential will be of the form
$V(Q) \propto Q^{- 2 p}$.  If the K\"ahler function is just given by
$K = Q \bar Q$, then the kinetic terms are standard, and we recover
the SUGRA quintessence potential already studied in Ref.~\cite{bm_99}
\begin{equation}
\label{sugraVQnorm}
V(\tilde{Q})
 = e^{\kappa \tilde{Q}^2 / 2}
   \frac{\langle Z^2 \rangle ^2m_\PL^{2p}}{\tilde{Q}^{2 p}}.
\end{equation}
The physical consequences of the SUGRA corrections are numerous, and
the potential given by Eq.~(\ref{sugraVQnorm}) was studied in detail
in Refs.~\cite{bm_99,bm_99_2,mbr_00}. There, it was shown that these
corrections lead to a better agreement with the currently available
data. In particular, the equation of state parameter is now given by
$\omega_\QUINT \simeq -0.82$, a value closest to $- 1$ than in the
usual quintessence models. The calculation of the CMB multipoles in
presence of SUGRA quintessence also show that the theoretical
predictions are consistent with the most recent data, in particular
the MAXIMA-1 data~\cite{mbr_00,hwang}. On the other hand, it is clear
that we have assumed that the $Q$ kinetic terms are canonical. If this
hypothesis is not fulfilled, potential~(\ref{sugraVQnorm}) is modified
and we see that the form of the potential above strongly depends on
the K\"ahler potential.
\par
\begin{figure}
\centerline{ 
\psfig{file=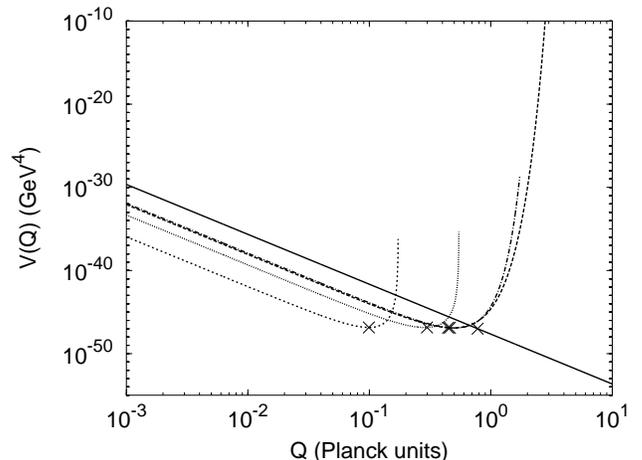,angle=270,width=3.5in}}
\caption{Comparison between Ratra-Peebles and SUGRA potentials. 
  The Ratra-Peebles potential [Eq.~(\ref{potRP}), solid line] is
  simply an inverse power law and always decreases. The standard SUGRA
  potential [Eq.~(\ref{sugraVQnorm}), long-dashed line] possesses an
  exponential correction which dominates when the field takes values
  close to the Planck mass. The other SUGRA potential we have
  considered in Eq.~(\ref{eq_vq_a}) is plotted for $a = 100$
  (short-dashed line), $a = 10$ (dotted line), and $a = - 1$
  (dot-dashed line).  All the curves were plotted with $\alpha = 2 p =
  6$ and normalized so that the quintessence field has a density
  parameter $\Omega_\QUINT = 0.7$ today, which roughly corresponds to
  put the minimum of the potential at $\rho_\CRIT$.  In addition, with
  crosses we have indicated the position of the quintessence field
  today.  It is clear that the field has almost reached the minimum of
  its potential in all (SUGRA) cases.}
\label{fig_pot}
\end{figure}

Let us study how the scalar potential is modified when more terms in
the K\"ahler potential are taken into account. In particular, one
would like to know whether the observables (for example, the equation
of state parameter) are strongly dependent on the higher terms in
series~(\ref{KQ}). Therefore, in order to have a more accurate
description of the true K\"ahler potential, it is interesting to take
into account one more term, and to choose
\begin{equation}
K = |Q|^2 + a \frac{|Q|^4}{m_\PL^2},
\end{equation}
where $a$ is a new free parameter, leading to the following exact
function $f(Q)$
\begin{eqnarray}
\label{fexact}
\tilde{Q} &=& \frac{1}{\sqrt{2}}\biggl[Q\sqrt{1+4a\frac{Q^2}{m_\PL^2}}
\nonumber \\
& & +\frac{m_\PL}{2\sqrt{a}}\ln \biggl(2\sqrt{a}\frac{Q}{m_\PL}
+\sqrt{1+4a\frac{Q^2}{m_\PL^2}}\biggr)\biggr].
\end{eqnarray}
Unfortunately, this function cannot be inverted exactly. However for
our purpose, it is sufficient to find the corrected potential at
leading order in the expansion in $Q/m_\PL$. One finds
\begin{equation}
\label{eq_vq_a}
V(Q) = \frac{M^{4 + 2 p}}
            {\left(Q - a \frac{Q^3}{3 m^2_\PL}\right)^{2 p}}
       \exp \left[\frac{\kappa}{2} \left(Q^2 - a \frac{Q^4}{6
m_\PL^2}\right)
            \right].
\end{equation}
Some examples of this potential are plotted in Fig.~\ref{fig_pot}.
\par
\begin{figure}
\centerline{
\psfig{file=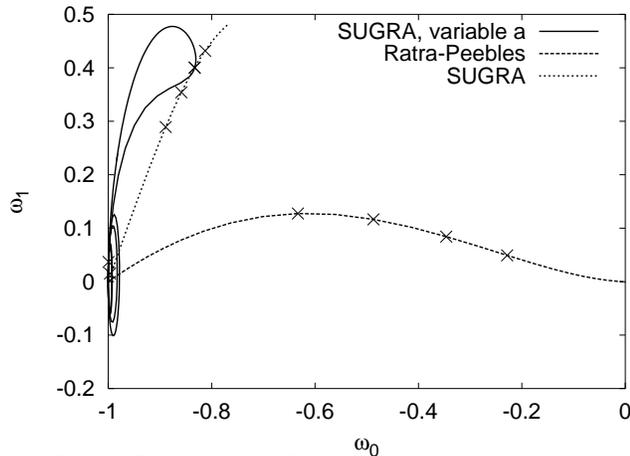,angle=270,width=3.5in}}
\caption{Effect of a modification to the quintessence
  potential [Eq.~(\ref{eq_vq_a})] on today's evolution of the
  quintessence equation of state (solid line). The three values of $a$
  plotted in Fig.~\ref{fig_pot} are represented with crosses (the case
  $a = - 1$ is near the intersection with the short-dashed line and
  the two others are near $\omega_0 \to -1$, $\omega_1 \to 0$).  As
  explained in the text, almost any value of $a$ helps the
  quintessence field to mimic a cosmological constant ($\omega_0 \to
  -1$, $\omega_1 \to 0$).  In addition, we have plotted the dependence
  on $\alpha$ of the Ratra-Peebles (long-dashed line) and SUGRA
  (short-dashed line) potentials. For the two curves, $\alpha$ varies
  from $ \ll 1$ (left) to $\gg 1$ (right).  The fact that the field
  roughly behaves as a cosmological constant for low values of
  $\alpha$ comes from the fact that the potential is flatter and,
  therefore the field stops more rapidly when it begins to dominate
  (even in the tracking regime, $1 + \omega_\QUINT \propto \alpha$,
  see Ref.~{\protect\cite{SWZ}}). Conversely, for high values of
  $\alpha$, the field tends to mimic the behavior of the background
  fluids. On these two curves, values of $\alpha = 2$, $4$, $8$, and
  $16$ are represented with crosses. The dependence on $\alpha$ of the
  SUGRA potential is much less important than in the Ratra-Peebles
  case.}
\label{fig_qa}
\end{figure}

Let us now study how the corrections described above can affect the
global evolution. In particular, as mentioned above, one would like to
know whether observable quantities are significally modified by the
new terms that we have considered in the series defining the K\"ahler
potential. An interesting way to distinguish between these various
models observationally is to look at the behavior of the quintessence
field equation of state. It can be shown~\cite{astier} that, provided
one knows both the matter density of the Universe as well as its
curvature, one can both recover the quintessence equation of state
parameter $\omega_\QUINT$, as well as its derivative today by studying
the luminosity distance vs redshift relation, for example with
supernovae type Ia. The parameter $\omega_\QUINT$ can be approximated
at low redshift by
\begin{equation}
\label{omega01}
\omega_\QUINT \simeq \omega_0 + z \omega_1 ,
\end{equation}
and both $\omega_0$ and $\omega_1$ can be recovered, at least in
principle, with good data.  In the case described by
Eq.~(\ref{eq_vq_a}), for positive values of $a$, the potential has a
steeper part around $Q = \sqrt{3 / a} m_\PL$ because it diverges.
Therefore the potential possesses a minimum before $Q = \sqrt{3 / a}
m_\PL$. As already stated, the field has reached the usual tracking
regime at earlier time (which corresponds to small values of $Q$);
therefore, it reaches its minimum sooner in the case of large $a$.  As
a consequence, the quintessence field behaves more rapidly as a
cosmological constant than in the standard SUGRA case, and of course
than in the Ratra-Peebles case.  This can be seen explicitly by
looking at the position of the quintessence field on its potential
(see Fig.~\ref{fig_pot}) or by plotting $\omega_\QUINT$ and its
derivative today as a function of $a$ (see Fig.~\ref{fig_qa}). Note,
however, that, strictly speaking, at the end of the evolution, all the
terms in the expansion of $f^{-1}(Q)$ should be taken into account
since $Q/m_\PL \simeq 1$. Therefore, the present calculation can only
give a hint of what happens when the corrections in the K\"ahler
potential are fully considered. For negative values of $a$, the
potential does not diverge but grow faster because of the higher
argument of the exponential part. Therefore, as for the $a > 0$ case,
the minimum of the potential occurs at lower values of $Q$, and the
field behaves more rapidly like a cosmological constant.

The main conclusions that we can draw from the previous analysis are
the following. In the context of effective SUGRA, there exists a
K\"ahler function and a superpotential which lead to a class of model
described by Eq.~(\ref{sugraVQ}). However, these models depend on
specific assumptions for the superpotential and K\"ahler functions. If
more generic terms are considered in the series defining the K\"ahler
potential, then some sensitivity of the observables to the form of the
K\"ahler potential within this class of models is found, but as long
as the potential possesses a minimum around $Q \simeq m_\PL$, the main
features of the SUGRA potential of Refs.~\cite{bm_99,bm_99_2} are
preserved.  Having identified the main advantages and drawbacks of the
approach followed in Refs.~\cite{bm_99,bm_99_2}, we now turn to a
different method where some of the previous shortcomings can be
avoided.

\section{A generic approach to quintessence with two scales}
\label{sec_gen}

In this section, which constitutes the core of this paper, we adopt a
different approach compared to that of Sec.~\ref{sec_RP}.  Since {\it
  a priori} there is no reason to consider that $m_\STRING$ and
$m_\PL$ are of the same order of magnitude, we do not make this
artificial assumption. As a consequence, we consider that $m_\STRING$
can have any value provided, of course, that it is smaller than the
Planck mass, $m_\STRING \ll m_\PL$.  Then, the strategy is as follows:
instead of trying to find the K\"ahler potential and the
superpotential which leads to the Ratra-Peebles potential as in
Refs.~\cite{bm_99,bm_99_2}, we will try to see which kind of potential
arises from a generic K\"ahler potential and superpotential, i.e.,
without any fine-tuning of their shape. We still assume that there are
three sectors in the theory, two of them being hidden. We first
investigate this question in the context of global SUSY, i.e., when
the value of the quintessence field is small in comparison to the
Planck mass, which is the case just after reheating where the initial
conditions are set. We assume that the K\"ahler potential is a
nonsingular series as $Q$ goes to zero. Let us expand the K\"ahler
potential focusing on the coupling between the quintessence field $Q$
and $Y$, the Polonyi field in the quintessence sector. One has
\begin{equation}
\label{conds}
K(Y, Q, \dots)
 =   \vert Y \vert^2
   + \vert Y \vert^2 \sum_{p = 1}^{p_\MAX}
                     \frac{1}{m_\STRING^{2 p}} (Q \bar Q)^p
   + \hat K (\dots),
\end{equation}
where $\hat K$ parametrizes the rest of the expansion (but of course
needs not to be equal to the one introduced previously). This
expression should be compared with Eqs.~(\ref{KWplb})
and~(\ref{KWplb2}).  This time the term $|Y|^2$ is present since we
have not assumed anything about the series defining the K\"ahler
potential. The key point is that we have only included terms sensitive
to the string scale and not the Planck scale because, in the limit of
global SUSY, this one is sent to infinity and therefore the
corresponding terms vanish. We have only assumed that the series can
be expressed as a polynomial. If this is not the case then a whole
knowledge of nonperturbative string theory is required. However,
truncating the whole series at the order $p_\MAX$ would require a
dynamical explanation which cannot be provided unless in a particular
model. For this reason we will study the dependence of the physical
observables on the degree of the polynomial.
\par
\begin{figure}
\centerline{
\psfig{file=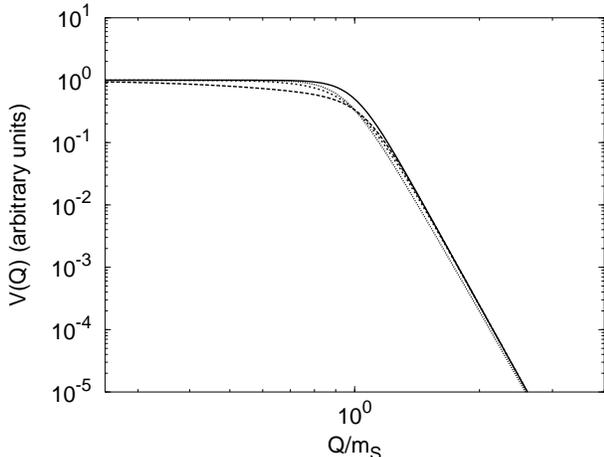,angle=270,width=3.5in}}
\caption{Different examples of potentials given by
  Eq.~(\ref{potSUSY}). The four curves represent potentials $V(Q) = |
  F_Y |^2 / [1 + (Q/m_\STRING)^{p_\MAX}]$ (solid line) and $V(Q) = |
  F_Y |^2 / [1 + (Q / m_\STRING)^{p_\MIN} + (Q/m_\STRING)^{p_\MAX}]$
  with $p_\MAX = 12$ and $p_\MIN = 2$ (long-dashed line), $p_\MIN = 6$
  (short-dashed line) and $p_\MIN = 10$ (dotted line).}
\label{fig_qa7}
\end{figure}

Let us calculate the corresponding scalar potential (assuming that the
quintessence field is real). The only term coming from the K\"ahler 
function which gives a contribution to the potential is given 
by
\begin{equation}
\label{Ka}
K^{\bar{Y} Y} = \frac{1}{\displaystyle
                         1 + \sum_{p = 1}^{p_\MAX} Q^{2p}/m_\STRING^{2
p}},
\end{equation}
from which we deduce that
\begin{equation}
\label{potSUSY}
V(Q) = \frac{\vert F_Y \vert^2}
            {\displaystyle 
             1 + \sum_{p = 1}^{p_\MAX} Q^{2 p} / m_\STRING^{2 p}}.
\end{equation}
Let us study this class of potentials in more detail. Typically, they
have the shape represented in Fig.~\ref{fig_qa7}. Whatever the precise
form of the series, for values of the field such that $Q \ll
m_\STRING$ the potential is almost flat since the constant term $1$
dominates in Eq.~(\ref{potSUSY}). This means that we no longer have a
divergence of the potential at small $Q$. When the field becomes of
the order of the string scale, $Q \simeq m_\STRING$, the precise form
of the series matters. But this is true only in a limited region, and
one expects that this will not affect the global behavior of
quintessence. In the region where $Q \gg m_\STRING$, only the term $(Q
/ m_\STRING)^{2 p_\MAX}$ is important, and the potential reduces to
\begin{equation}
\label{limitpot2s}
V(Q) \simeq \frac{\vert F_Y \vert^2 m_\STRING^{2 p_\MAX}}{Q^{2 p_\MAX}},
\end{equation}
i.e., we recover a Ratra-Peebles potential, and again the detailed
form of the series does not matter. This region corresponds to the
straight line in Fig.~\ref{fig_qa7} (the slope of which is $- 12$
since $p_\MAX = 6$ was chosen for this plot). Since the tracking
behavior essentially depends on the behavior of the field at {\em late
  times}, i.e., before it reaches the Planck mass, the attractor
properties of the standard Ratra-Peebles potential still hold in this
case, as we have checked numerically. Therefore, the modifications in
the potential at low $Q$ do not matter as long as $m_\STRING$ is not
too large (typically, $m_\STRING$ must be $\sim 2$ orders of magnitude
smaller than the Planck mass; this bound is even relaxed for high
values of $\alpha$). Such a behavior was already remarked upon in
another context when we considered the quantum corrections to the
Ratra-Peebles potential~\cite{bm_99_2}. Note, however, the explicit
dependence on the degree $p_{\MAX}$. This has important observable
consequences. Indeed, it appears that potential~(\ref{limitpot2s})
leads to an equation of state $\omega_0$ which exhibits a strong
dependence on $p_\MAX$.  This is less true for the derivative of the
equation of state $\omega_1$, as can be seen from Fig.~\ref{fig_qa}
(dashed line).  However we shall see that this problem is far less
serious when SUGRA corrections are included, in which case the values
of $(\omega_0, \omega_1)$ accumulate numerically around $(-0.8, 0.45)$
in the large $p_\MAX$ regime (Fig.~\ref{fig_qa}, dotted line). This is
an interesting indication that the physical observables are stable
with respect to variations of the truncation degree.
\par
Another important consequence is that the SUSY breaking scale is now
given by
\begin{equation}
\label{SUSYbreak}
\vert F_Y \vert^2 
\simeq \rho_\CRIT \left(\frac{m_\PL}{m_\STRING}\right)^\alpha ,
\end{equation}
where $\alpha \equiv 2p_\MAX$. For $m_\PL = m_\STRING$ one recovers
the usual result given in Eq.~(\ref{breakmod}). However, the important
point is that in the present framework, $m_\PL$ and $m_\STRING$ do not
need to be the same, which has the important consequence that now the
SUSY breaking scale in the quintessence sector decouples from the
critical energy density.  Let us show a few orders of magnitude. In
particular, one would like to fix the SUSY breaking scale in the
quintessence sector to the same value as the SUSY breaking scale in
the broken sector, i.e., $\langle F_P\rangle \simeq 10^{20}
\UUNIT{GeV}{}$. This would be one step toward an identification
between the quintessence sector and the broken sector, thus leaving
only one hidden sector. This strategy will be pursued in
Sec.~\ref{sec_SUSY}. Fixing $\vert F_Y \vert \simeq 10^{20}
\UUNIT{GeV}{2}$ and writing $m_\STRING = 10^{-x} m_\PL$, we find that
$x \simeq 67/ \alpha$. We see that the string scale varies between the
$\UNIT{TeV}{}$ scale and the Planck mass for $\alpha > 3$. It is
auspicuous that to maintain a low value of the SUSY breaking scale in
the quintessence sector in the large $\alpha$ limit, we need to take
values of $m_\STRING$ which are closer and closer to the Planck scale.
As already stated, as long as $m_\STRING$ is a few order of magnitude
smaller than the Planck mass, this has no significant influence of the
evolution of the quintessence field today. The previous results follow
from the direct coupling between one field $Q$ and the SUSY breaking
field $Y$, and does not require any fine-tuning. In particular, the
presence of an inverse power law only requires that one can trust the
perturbative expansion of the Lagrangian, i.e., one does not need to
know the whole power series.
\par
We now need to take into account the SUGRA corrections. As in
Sec.~\ref{sec_RP}, the form of the potential is given by the positive
definite expression $V = e^{\kappa K} K^{Y \bar Y} \langle
\partial_YW\rangle ^2$, where again we have used the fact that the $D$
terms are not modified in SUGRA. The K\"ahler potential evaluated at
the minimum of the potential for the matter fields is a series in $1 /
m_\STRING$, and reads
\begin{equation}
\label{KQms}
K (Q, \left<\varphi^{i k}\right>)
 = \sum_{n = 1}^{n_\MAX} \frac{c_{2 n}}{m_\STRING^{2(n-1)}}(Q \bar Q)^n ,
\end{equation}
where we have fixed the other hidden sector fields to their vacuum
expectation values. This equation is similar to Eq.~(\ref{KQ}). Note,
however, that we have only kept the dominant $1 / m_\STRING$ terms. If
$m_\STRING \simeq m_\PL$ we only need to substitute $m_\PL$ for
$m_\STRING$ in the expansion. The kinetic term of $Q$ is not
normalized. To deal with this problem, as previously, we define a new
scalar field according to Eq.~(\ref{staQ}) (of course, now, the
function $f$ needs not to be the same). This leads to the potential
\begin{equation}
V(\tilde Q)
 = e^{\kappa K[f^{- 1} (\tilde Q)]}
   \frac{\langle \partial_YW\rangle ^2}
        {\displaystyle
         1 + \sum_{p = 1}^{p_\MAX} [f^{- 1} (\tilde Q)]^{2 p} 
                                          / m_\STRING^{2 p}}.
\end{equation}
The previous equation gives the generic prediction for any theory
which can be effectively described by SUGRA with two scales.  Note
that taking $m_\PL \to \infty$, this reduces to the globally
supersymmetric result, as expected. Now we can deduce the form of the
potential in the three different regimes, and study how it is affected
by the particular form of the theory.  First, we note that it does not
depend on the superpotential: it is sufficient to have $\langle
\partial_YW\rangle \neq 0$, i.e., a Polonyi field in the quintessence
sector. When $Q \ll m_\STRING$ then $\tilde Q = \sqrt 2 Q$ and
$V(\tilde Q) \simeq \langle \partial_YW\rangle ^2$. The potential no
longer blows up. In this regime, it does not depend on the details of
series~(\ref{conds}) or~(\ref{KQms}). For $Q \simeq m_\STRING$ all the
terms in the expansion play a role, and the precise shape of the
potential cannot be determined unless a specific model is given. But
again we expect that we will not affect the cosmological observables
since they are determined in a regime where $Q = m_\PL \gg m_\STRING$.
For large $Q$ the highest power is only required. As we are interested
in the $Q \simeq m_\PL$ regime we conclude that
\begin{eqnarray}
K_{Q \bar Q}
 & = & n_\MAX^2 c_{2 n_\MAX}
       \frac{(Q \bar Q)^{n_\MAX - 1}}{m_\STRING^{2(n_\MAX - 1)}} \\
\Rightarrow
\tilde{Q}
 & = & \sqrt{2 c_{2 n_\MAX}} \frac{Q^{n_\MAX}}{m_\STRING^{n_\MAX - 1}},
\end{eqnarray}
leading to
\begin{equation}
\label{genepot}
V(\tilde Q)
 = A e^{\kappa\frac{\tilde Q^2}{2}}
   \frac{\langle \partial_YW\rangle ^2}{\tilde Q^{2 p_\MAX / n_\MAX}} ,
\end{equation}
where $A = (\sqrt{2 c_{n_\MAX}} m_\STRING)^{2 p_\MAX / n_\MAX}$.  Note
that the coefficients arrange themselves such that $\kappa
\tilde{Q}^2/2$ appears in the potential without any additional
multiplicative factor in the argument of the exponential. We can
identify $\alpha \equiv 2 p_\MAX /n_\MAX$. Therefore, in this regime,
we recover the SUGRA quintessence potential, which now appears as a
generic property of any effective SUGRA theory with two scales.  Now
the degrees of the truncated series $n_\MAX $ and $p_\MAX$ play
competing roles. In particular, three natural behaviors can occur. In
the first one $\alpha$ goes to zero. This is physically disfavored, as
this would require that $\langle \partial_Y W\rangle ^2$ converge to
the critical energy density $\rho_\CRIT$. Similarly $\alpha$ can go to
infinity, with the need for $m_\STRING$ to be closer and closer to the
Planck scale.  Finally, $\alpha$ can remain finite. In this case we do
not need to fine-tune the SUSY breaking scale. The point is that the
observables do not depend very much on $\alpha \equiv 2 p_\MAX
/n_\MAX$. Indeed, a large range of values of $\alpha$ lead to the same
CMB spectrum and the same dependence of the equation of state at small
redshifts (see Fig.~\ref{fig_qa}, dotted line and Ref.~\cite{mbr_00}).
It is remarkable that from an {\it a priori} very complicated theory,
we end up with the conclusion that observables like $(\omega_0,
\omega_1)$ are uniquely determined by potential~(\ref{genepot}).
Since typically, we expect that the coefficient $c_{n_\MAX}$ is of
order 1, we deduce that the SUSY breaking scale is again given by
relation~(\ref{SUSYbreak}). In order to justify that the previous
considerations really lead to a successful and realistic model for
quintessence, we need to study the process of SUSY breaking in more
detail.

\section{Supersymmetry breaking}
\label{sec_SUSY}

In the previous sections we have seen that it is necessary to assume
three different sectors, two of them being hidden. In this section, we
thoroughly analyze the consequences of SUSY breaking, both from
cosmological and particle physics points of view.

\subsection{Spontaneous vs explicit supersymmetry breaking}

A first study of SUSY breaking in the context of quintessence was made
in an interesting paper by Kolda and Lyth~\cite{kolda_lyth}.  There,
the authors pinpointed a possible incompatibility between quintessence
and SUSY. Indeed the expansion of Eq.~(\ref{SUSYL}) comprises the two
terms
\begin{equation}
\label{F}
K_{Y \bar Y}\vert F^Y \vert^2 + W_Y F^Y + \bar W_{\bar Y} F^{\bar Y}.
\end{equation}
Assuming that SUSY is broken {\it explicitly} by $F_Y $ leads to a
polynomial expansion of the scalar potential in $Q$ when using the
general Taylor expansion of $K_{Y \bar Y}$. Fortunately, in SUGRA one
must consider SUSY as a local gauge theory wich cannot be broken
explicitly, as the electroweak symmetry which is not broken by putting
an explicit gauge symmetry breaking mass in the Lagrangian.  SUSY is
broken {\it spontaneously} by the nonvanishing vev of $F$ terms
obtained by solving the equations of motion. This leads to a
super-Higgs mechanism, where the would-be massless Goldstone fermion
is eaten by the gravitino which becomes massive~\cite{supehig}. As the
$F$ terms are auxiliary field terms with no kinetic terms, one can
solve Eq.~(\ref{F}) algebraically to give
\begin{equation}
F^Y = - K^{Y \bar Y} \bar W_{\bar Y} ,
\end{equation}
leading to the potential investigated in the previous sections. It is
apt that an intrinsic feature of SUGRA prevents this type of
quintessential difficulty.

\subsection{Moduli supersymmetry breaking}

We have seen that a quintessence potential can be obtained in a hidden
(quintessence) sector. On the other hand, we have assumed that SUSY
was broken in another hidden sector. Therefore, one may wonder whether
it would not be possible to consider only one hidden sector where SUSY
is broken and, at the same time, to which the quintessence field
belongs. In this section, we will include the effects due to other $F$
terms, and study the modifications that they impose on the potential.
In particular we suppose that these are the single K\"ahler moduli $T$
and the dilaton field $S$ where the superfields $T$ and $S$ belongs to
the unique (postulated) ``broken-quintessence'' hidden sector. Because
of SUSY breaking the potential will have the form
\begin{eqnarray}
V_\BROKEN (Q)
 & = & V(Q)+e^{\kappa K}\biggl( K^{T \bar T} |F_T|^2
       + K^{S \bar S} |F_S|^2\biggr)
\nonumber \\ & &
       + |D|^2
       - 3 m^2_{3 / 2} / \kappa 
       +V_\ADD .
\label{VB}
\end{eqnarray}
where the potential $V(Q)$ is the quintessence potential obtained
previously. The $D$ terms are independent of $Q$, as this is a neutral
field. The gravitino mass $m_{3 / 2}$ is nonzero due to the breaking
of SUSY. The last term $V_\ADD$ springs from the visible sector and
gives large contributions to the cosmological constant. This is the
cosmological constant problem: $V_\BROKEN (Q)$ contains huge constant
terms which, {\it a priori}, dominate all the other contributions.
The $F_S$ and $F_T$ auxiliary fields are given by
\begin{equation}
F_{S,T}=-\partial_{S,T}W-\kappa(\partial_{S,T}K)\ W ,
\end{equation}
and depend on the nonperturbative corrections to the superpotential
which are responsible for the breaking of SUSY. There is a strong
dependence of $F_T$ and $F_S$ on the K\"ahler potential.  To go
further we need to return to Eq.~(\ref{conds}) and to be more specific
about the forms of the function $\hat{K}$. We take a generic form of
the K\"ahler function as
\begin{eqnarray}
\hat{K} 
 &=& \frac{1}{\kappa}  \left[ - 3 \ln(T + \bar T) - \ln(S + \bar
S)\right]
\nonumber \\
 &   & + m_\STRING^2 \sum_{p q k} d_{p q  k} 
(S + \bar S)^{- p} (T + \bar T)^{-q}
\biggl(\frac{Q \bar Q}{m_\STRING^2}\biggr)^k ,
\label{KK}
\end{eqnarray}
where we assume that this is a polynomial in $Q \bar Q$ (the
coefficients $d_{p q k}$ are just the coefficients of the polynomial).
Only inverse powers of $m_\STRING$ were taken into account, as order
by order in $Q$ the inverse powers of $m_\PL$ are suppressed.
Computing the derivative with respect to $S$ and $T$ leads to a
polynomial dependence on $Q$ of $F_S$ and $F_T$. This implies that the
SUSY breaking scale varies during the evolution of the universe, and
therefore the sparticle masses become strongly time dependent. Indeed,
the mass matrix of the scalars depends explicitly on the $F$ terms,
\begin{equation}
m^2_{A\bar B}=e^{\kappa K}\biggl(\frac{\kappa}{3}K_{A\bar B}K_{C\bar
  D}-R_{A\bar BC\bar D}\biggr)F^C\bar F^{\bar D},
\end{equation}
where the second term involves the Riemann tensor deduced from the
K\"ahler potential. It is easy to see that a polynomial dependence on
$Q$ for $F_S$ and $F_T$ leads to a polynomial dependence on $Q$ of the
masses from $K^{T\bar T}F_T \bar F_{\bar T}$ and $K^{S\bar S}F_S \bar
F_{\bar S}$. At large $Q$ this behaves like $(Q/m_\STRING)^{2k_\MAX}$,
where $k_\MAX$ is the dominant term in Eq.~(\ref{KK}).
\par 
To avoid this we must conclude that the quintessence field decouples
from the SUSY breaking sector
\begin{equation}
d_{p q k}=0,\quad k\ne 0 .
\end{equation}
On the whole we find that the SUSY breaking sector and the
quintessence sector must be separate.
\par
Coming back to Eq.~(\ref{VB}), there is a negative contribution from
the gravitino mass:
\begin{equation}
m_{3 / 2} = \kappa e^{\kappa K/2}\langle W \rangle .
\label{m32}
\end{equation}
Combining with the $F_S$ and $F_T$ terms, this leads to the following
term in the potential $V_\BROKEN$:
\begin{equation}  
e^{\kappa K}\biggl( K^{T \bar T} |F_T|^2 + K^{S \bar S} |F_S|^2 - 3
\kappa\langle W\rangle ^2\biggr).
\end{equation}
In the early universe this is a cosmological constant as the term in
brackets is a constant. As $Q$ increases the exponential corrections
become relevant. So this term acts as a slowly varying cosmological
constant.  Moreover, we can expect a large contribution $V_\ADD \simeq
m_{\rm W}^4$ from the visible sector. Both contributions should be
large compared to the critical density of the universe.  Nevertheless,
there is a strong constraint springing from the existence of an
attractor. The attractor condition [Eq.~(\ref{atra})] should be
compatible with the requirement that the total potential reproduces
$\Omega_\Lambda \rho_c$. It can easily be seen that, if the slowly
varying and constant contributions are much larger than the critical
density, then the attractor disappears. Consequently we shall assume
that the extra constant and slowly varying pieces in the potential
vanish altogether. This is another manifestation of the fact that it
is necessary to assume that the cosmological constant problem is
solved before considering the quintessence hypothesis. In the context
of quintessence, the relevant question is whether the {\it dynamical}
part of the potential after SUSY breaking is modified.  In particular,
this leads to the requirement that the contributions from the visible
sector and the broken sector must vanish independently, i.e.,
\begin{equation}
V_\ADD=0, \quad K^{T \bar T} |F_T|^2 + K^{S \bar S} |F_S|^2=3 
\kappa\langle W\rangle ^2 .
\end{equation}
The second of these constraints is the usual fine-tuning of the SUSY 
breaking sector.
\par
Let us now consider the contribution to the scalar masses due to the
Polonyi field $Y$,
\begin{equation}
\biggl(\frac{\kappa}{3}K_{A\bar B}- R_{A\bar B Y\bar Y}K^{Y\bar Y}\biggr)V(Q),
\end{equation}
which is negligible now due to the smallness of $V(Q)$. The scalars
receive a mass from the $F_S$ and $F_T$ terms, which
reads~\cite{pes,lou}
\begin{eqnarray}
m^2_{A\bar B} &=& m^2_{3/2}K_{A\bar B} \nonumber \\
& & -e^{\kappa K}(R_{A\bar B S\bar S}F^S\bar F^{\bar S} +
R_{A\bar B T\bar T}F^T\bar F^{\bar T}).
\end{eqnarray}
Note that the sparticle masses will have a universal redshift
dependence coming from exponential factor in Eq.~(\ref{m32}).  This
dependence is only relevant in the recent past. It would be
interesting to study the associated phenomenology. There is a final
constraint springing from the gauginos masses~\cite{pes,lou}
\begin{equation}
m_a=\frac{\sqrt \kappa}{2}e^{\kappa K} F^I\partial_I\ln g_a^{-2},
\end{equation}
where $g_a$ is the gauge coupling of the $a$th gauge group.  To
leading order one can expand
\begin{equation}
g_a^{-2}= S+\bar S +\beta \sqrt\kappa (Y+\bar Y),
\end{equation}
where we have included a dependence on $Y$. This is what happens in
type I models if the Polonyi field can be identified with the blowing
up moduli.  Nevertheless the presence of $K^{Y\bar Y}$ implies that
the $F_Y$ contribution is negligible.  So we find that that the masses
of sparticles do not depend on $F_Y$. This allows for independent
supersymmetry mechanisms in the ``broken'' and ``quintessence''
sectors. In particular the mechanism of Sec.~\ref{sec_RP}, where
\begin{equation}
F_Y^2 = \langle Z^2 \rangle^2
      = \biggl(\frac{m_\PL}{m_\STRING}\biggr)^{\alpha} \rho_\CRIT ,
\end{equation}
is viable. Phenomenologically we should impose that the corresponding
Fayet-Iliopoulos term is larger than the weak scale.  This leads to
\begin{equation}
\frac{m_\STRING}{m_\PL}
 \le \biggl(\frac{\rho_\CRIT}{m_W^4}\biggr)^{1/\alpha},
\end{equation}
which is reasonable as soon as $\alpha>3$. We can even go further by 
noticing that the Fayet-Iliopoulos term is of the
order of the string scale. Imposing that $F_Y=m_\STRING^2$ leads to
\begin{equation}
\rho_\CRIT=\frac{m_\STRING^{4+\alpha}}{m_\PL^{\alpha}}.
\end{equation}
In new type I string scenarios the string scale can be as low as 
the TeV region. In that case this leads to $\alpha=4$. This 
determines $p_\MAX=2$ for a flat K\"ahler potential in 
$Q$. The relation 
\begin{equation}
\rho_\CRIT^{1/4}=\frac{m_\STRING^2}{m_\PL}
\end{equation}
was advocated in Ref.~\cite{hall} to obtain a natural solution to the
coincidence problem. We find that it can be embedded in a SUGRA
description of quintessence with two scales.
\par
In conclusion we have seen that quintessence is compatible with SUSY
breaking, and should belong to a hidden sector different from the
hidden broken sector.

\section{Influence of a minimum in the quintessence potential}
\label{sec_min}

From the above considerations, it seems that a generic consequences of
taking into account high energy physics is the presence of a minimum
in the quintessence potential. This differs from the Ratra-Peebles
case, where the potential is monotonic and goes to zero at infinity.
Therefore, one may wonder what the physical consequences of the
presence of this minimum are. The purpose of this section is to study
this question.

\subsection{Oscillations of the quintessence field}

\begin{figure}
\centerline{
\psfig{file=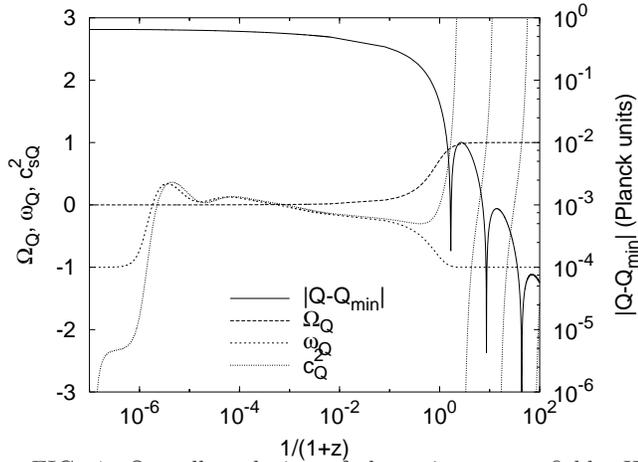,angle=270,width=3.5in}}
\caption{Overall evolution of the quintessence field. We start
  at a redshift of $z = 10^7$ with a quintessence field initially at
  rest ($\omega_\QUINT \to - 1$, short-dashed line) and subdominant
  ($\Omega_\QUINT \to 0$, long-dashed line). Then the field joins the
  attractor around $z \simeq 10^4$. It remains on this attractor as
  long as it is subdominant, i.e., $\Omega_\QUINT \ll 1$. When it
  starts to dominate, it gradually behaves as a cosmological constant
  ($\omega_\QUINT \to - 1$ again around $z \simeq 0$). Then the field
  experiences some damped oscillations around its minimum (solid
  line).  The behavior of the parameter $\omega_\QUINT$ can be studied
  by looking at the variable $c_\QUINT^2 \equiv \dot p_\QUINT / \dot
  \rho_\QUINT$ (dotted line), which diverges when $\omega_\QUINT$
  reaches $- 1$ [this occurs initially and when $Q(t)$ reaches an
  extremum].}
\label{fig_az}
\end{figure}

The SUGRA potential possesses a minimum located at $Q_\MIN =
\sqrt{\alpha} / \kappa^{1 / 2} = \OO(m_\PL)$, see
Fig.~\ref{fig_pot}. Thus, {\it a priori}, this could modify the final
evolution of the field. Therefore, let us expand the field around the
minimum; we write
\begin{equation}
\label{expanfield}
\bar{Q} = \sqrt{\alpha} + \bar{q},
\end{equation}
where $\bar{Q} \equiv \kappa^{1 / 2} Q$ is dimensionless and where
$\bar{q}$ is a small quantity. If we neglect the quadratic order, the
Einstein equation reads $H^2 = H_0^2 = (\kappa / 3) V(\sqrt{\alpha})$
which implies that $a(t) = a_0 e^{H t}$. On the other hand, the
Klein-Gordon equation is given by
\begin{equation}
\label{kgosci}
\ddot{\bar{q}} + 3 H_0 \dot{\bar{q}} + 6 H_0^2 \bar{q} = 0 .
\end{equation}
The solution to this equation is given by the following expression:
\begin{equation}
\label{sol}
\bar{q}(t) 
 \propto \exp \left[\left(- \frac{3}{2} 
                              \pm i \frac{\sqrt{15}}{2}\right) H_0
t\right] .
\end{equation}
This solution is oscillatory with a damping term proportional to $a^{-
  3 / 2}$. The period of the oscillations is equal to $ \simeq
H_0^{-1}$, i.e., is equal to the age of the Universe today. Therefore,
it is clear that no oscillation took place until now since the age of
the Universe is the time already necessary to reach the region where
the oscillations could occur. Conversely, the future of the Universe
will be different in comparison with the Ratra-Peebles potential case.
Numerically, for the case $\alpha = 11$, the redshifts at which the
field stops are $z=-0.65, -0.92, -0.98, \dots$ etc. The first redshift
corresponds to $a / a_0 \sim 2.85$, i.e., to a time where the scale
factor is $2.85$ larger than today (see Fig.~\ref{fig_az}). This is of
course independent of the initial conditions provided that we are
initially in the allowed range.

It is of course possible that some oscillations occur before today,
but this is not easy. The main reason is that the quintessence field
rolls rather slowly toward the bottom of its potential, so that the
quintessence density parameter $\Omega_\QUINT$ is almost equal to $1$
at the time where the field stops for the first time (as can be seen
in Fig.~\ref{fig_az}). Another possibility is that $\Omega_\QUINT$ is
of order unity at early time. In this case, the field is initially
very small, and correspondingly its energy density is large. Then, the
field is in a ``fast-roll'' regime, i.e., $\omega_\QUINT \simeq 1$,
and is not slowed down enough by the expansion. It then goes through
(still in a fast-roll regime) its minimum, and is stopped by the very
steep exponential growth of the potential at large $Q$. Such a
behavior does not affect the behavior of the quintessence field today,
but can leave some imprints in the high frequency part of the
primordial gravitational wave spectrum (see, e.g.,
Ref.~\cite{ur_2000}).

\subsection{Amplitude of the minimum}

Let us now study the influence of a pure cosmological constant term in
the quintessence potential. We would like to know whether we can
change the value of the minimum and, in particular, whether it is
possible to put it to zero. Therefore, we take a SUGRA potential to
which we add a constant term
\begin{equation}
V (Q)
 =   e^{\kappa Q^2 / 2} \frac{M^{4+\alpha}}{Q^\alpha}
   + (X_\MIN - 1) V_\MIN .
\end{equation}
In this expression $V_\MIN $ is the value of the potential at its
minimum, i.e., for $Q=\sqrt{\alpha /\kappa }$ and $X_\MIN \geq 0$ a
free parameter. $X_\MIN =0$ corresponds to a vanishing minimum, and
$X_\MIN = 1$ reduces the above potential to the standard SUGRA
potential. We would like to emphasize that there is no fine-tuning of
the location of the minimum, it follows directly from the shape of the
potential (and is of course independent of the constant $M$). The fact
that the field is today near the minimum of the potential follows
directly from the fact that, because of the presence of the attractor,
the field is today of the order of the Planck mass, which also turns
out to be the location of the minimum of the potential. Again, no
fine-tuning is required to have this property which arises naturally
in SUGRA quintessence.
\par
\begin{figure}
  \centerline{ \psfig{file=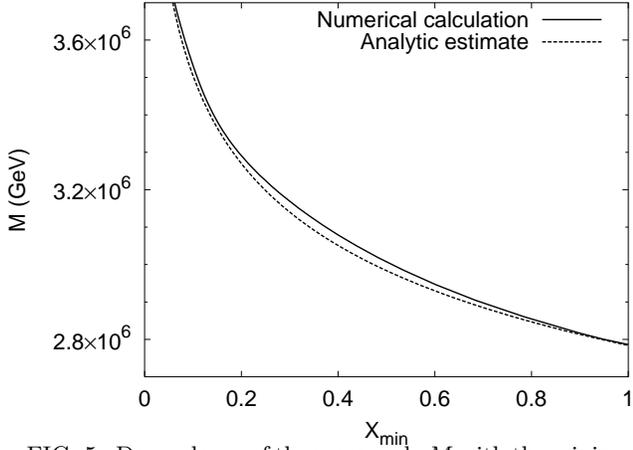,angle=270,width=3.5in}}
  \caption{Dependence of the mass scale $M$ with the minimum of the
  potential $X_\MIN$. The numerical computation gives a very good
  agreement with the estimate of Eq.~(\ref{M_val2}).}
\label{fig_qa9}
\end{figure}
\begin{figure}
  \centerline{\psfig{file=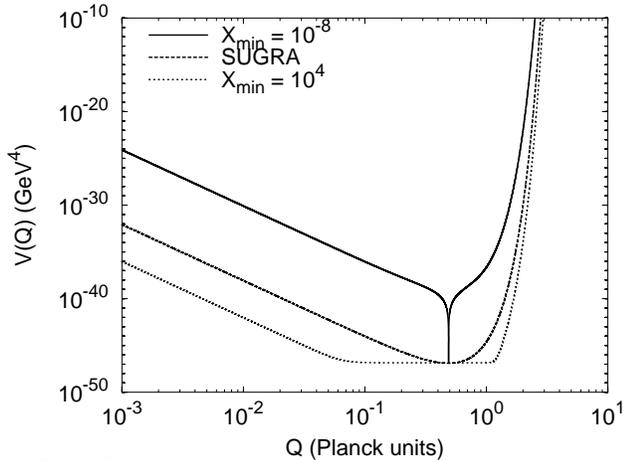,angle=270,width=3.5in}}
  \caption{Shape of the quintessence potential for various values of
    $X_\MIN$. The potentials were normalized so that $\Omega_\QUINT =
    0.7$ today, which roughly corresponds to requiring that the
    minimum of the potential is equal to the critical density today.
    Note the presence of a broad, flat region for high values of
    $X_\MIN$, and a deep and narrow depression for small values of
    $X_\MIN$. These features have a large importance on the evolution
    of the quintessence field today.}
\label{fig_qa8}
\end{figure}
\begin{figure}
\centerline{ 
\psfig{file=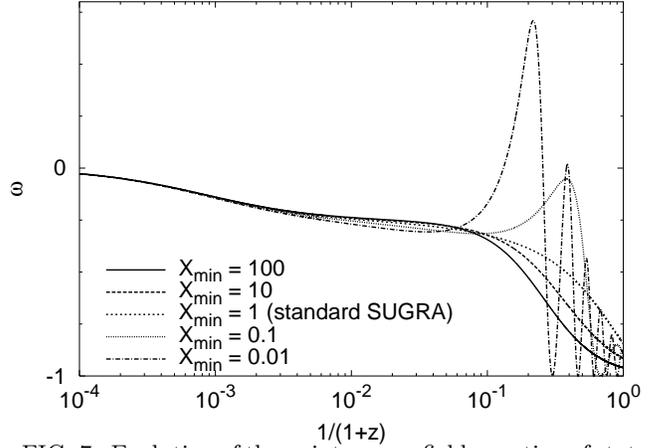,angle=270,width=3.5in}}
\caption{Evolution of the quintessence field equation of state
  parameter for several values of $X_\MIN$. The field starts at a
  moderately high redshift ($z \simeq 100$) from its attractor value
  (which means $\omega_\QUINT \simeq -0.25$ for $\alpha = 6$ here),
  and subsequently starts to behave as a cosmological constant as its
  energy density dominates (we have taken $\Omega_\QUINT = 0.7$
  today). As explained in the text, large values of $X_\MIN$ all lead
  to essentially the same beahavior, whereas low values of $X_\MIN$
  cause the field to oscillate.}
\label{fig_qa5}
\end{figure}
\begin{figure}
\centerline{
\psfig{file=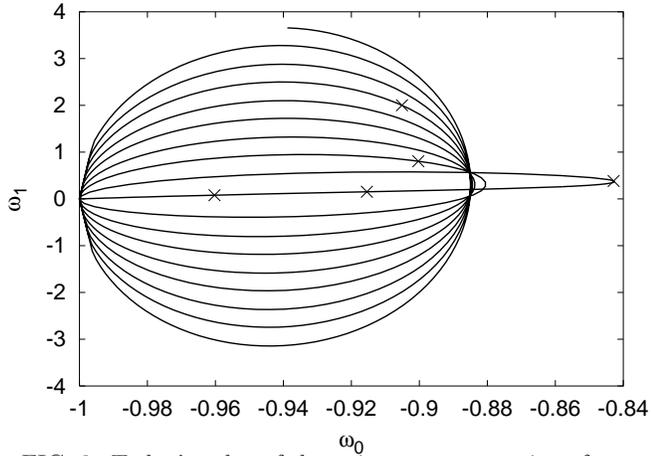,angle=270,width=3.5in}}
\caption{Today's value of the quintessence equation of state
  parameter $\omega_0$ and its derivative $\omega_1$ for several
  values of $X_\MIN$. The crosses represent the values of $X_\MIN$
  used in Fig.~\ref{fig_qa5} (same color code). Note that for low
  values of $X_\MIN$, the oscillations of the field are quite rapid,
  and therefore, the equation of state parameter is not very well
  approximated by $\omega \sim \omega_0 + z \omega_1$, even for a
  relatively short interval of the redshift.}
\label{fig_qa1}
\end{figure}

Let us start with the case where the minimum is not zero. The presence
of a constant term can influence the shape of the potential and the
value of the constant $M$, as explained below. Let us start with the
constant $M$. In all the cases presented here, as mentioned above, the
constant $M$ is found numerically by requiring that $\Omega_\QUINT =
0.7$ today. In all the cases of interest, the quintessence equation of
state is such that $- 1 \leq \omega_\QUINT
\leq 0$ today. This means that a significant part of the energy
density of the field is determined by its potential energy. In presence 
of the additional constant term, this implies that the constant $M$ is 
no longer given by Eq.~(\ref{M_val}), but rather by
\begin{equation}
\label{M_val2}
M \simeq
 \left(\frac{\rho_\CRIT m_\PL^\alpha}{X_\MIN}
 \right)^{\frac{1}{4 + \alpha}} .
\end{equation}
This is what we can check on Fig.~\ref{fig_qa9}.
Knowing how to determine $M$, we can now turn to the shape of the
potential. For large values of $X_\MIN$, there is a large region where
the potential is almost flat. This means that when the quintessence
field enters this region, it behaves very quickly as a cosmological
constant. Conversely, small values of $X_\MIN$ produce a deep and
narrow ``hole'' in the potential in which the field oscillates when it
falls into it. These two cases are represented in Fig.~\ref{fig_qa8}.
The addition of a constant term must have some observable consequences
today. This is what we can check in Fig.~\ref{fig_qa5}, where we plot
the value of $\omega_\QUINT$ as a function of the redshift. As
expected, large values of $X_\MIN$ do not significantly differ from
the $X_\MIN = 1$ case, except that the equation of state parameter
$\omega$ goes faster to $-1$ (the potential is less steep).
Conversely, the oscillations for small values of $X_\MIN$ are clearly
observable. This is due to the fact that in this case,
Eq.~(\ref{kgosci}) reads
\begin{equation}
\label{kgosci2}
\ddot{\bar{q}} + 3 H_0 \dot{\bar{q}} + 6 \frac{H_0^2}{X_\MIN} \bar{q} = 0 ,
\end{equation}
so that the frequency of the oscillations can be arbitrarily large.
Then if we plot the values of $(\omega_0, \omega_1)$ for several
values of $X_\MIN$, the oscillations of the field translate into
ellipses in the $(\omega_0, \omega_1)$ plane, see Fig.~\ref{fig_qa1}.

\begin{figure}
\centerline{
\psfig{file=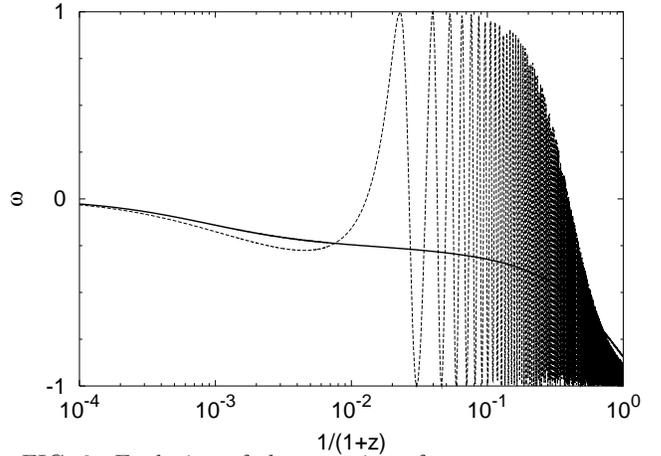,angle=270,width=3.5in}}
\caption{Evolution of the equation of state parameter
  $\omega_\QUINT$ as a function of the redshift for $X_\MIN = 1$
  (solid) and $X_\MIN = 10^{- 5}$ (dashed line). For such a low value
  of $X_\MIN$, only for a small redshift can one see the decay of
  $\omega_\QUINT$ towards $- 1$, as indicated by the decreasing
  envelope of the curve.}
\label{fig_qa6}
\end{figure}

Finally, we would like to stress some important properties of the
dynamic of the quintessence field in the case of a vanishing value of
$X_\MIN$, i.e., when one tries to set the potential to zero. A
decreasing value of $X_\MIN$ leads to an increasing number of
oscillations experienced by the quintessence field before today, see
Eq.~(\ref{kgosci2}). Numerically, this translates into a very weird
behavior of the function $\omega_\QUINT (z)$ as $X_\MIN$ goes to $0$,
see Fig.~\ref{fig_qa6}. Now, in the vicinity of the minimum of the
potential, the potential has a quadratic shape.  Therefore, this leads
to an equipartition between the kinetic energy and the potential
energy, and therefore to an average equation of state parameter
$\omega_\QUINT$ equal to $0$, a well-known behavior of the inflaton
field at the end of inflation~\cite{linde}\footnote{This point can in
  principle be evaded if we suppose that the potential behaves like
  $(Q - Q_\MIN)^\beta$, with $0 < \beta < 2$, but this seems to be an
  unlikely possibility in the case presented here.}. In this case, the
equation of state of the field is exactly the same as the one of
ordinary matter. As a consequence, the ratio between $\Omega_\QUINT$
and $\Omega_\MAT$ becomes a constant. This means that the value of
$\Omega_\QUINT$ today is approximatively given by the value of
$\Omega_\QUINT$ when the field started its oscillations, denoted
$\Omega_\QUINT^\OSC$ in what follows. Then, the relevant question is:
can $\Omega_\QUINT^\OSC$ be equal to (say) $0.7$?  The answer to this
question depends on the physical reason which causes the field to
leave the attractor. {\it A priori}, two situations can be envisaged.
First, the field leaves the attractor because it has not yet reached
(or felt) the minimun and it starts to dominate. This is what happens
in the Ratra-Peebles case (for which, anyway, there is no minimun).
Second, conversely, it has not yet started to dominate but the field
``feels'' the presence of a minimum. In the second case, by definition
we have $\Omega_\QUINT^\OSC \ll 1$, and the answer to the question
above is ``no''. Therefore, only the first situation remains a
possibility. Let us study this situation in more detail. In
particular, one may wonder whether it can really happen that the field
dominates before encountering the minimun. The field dominates when $Q
= Q_\END$, defined by the condition $\rho_\QUINT \simeq \rho_\MAT /
x$, where $x$ is an arbitrary number. A reasonable value for $x$ is,
for example, $x=10$. Using the equation of the attractor [see
Eq.~(\ref{attra})], it is easy to establish that $\kappa Q_\END^2
=\alpha (\alpha +2)/[3(x+1)]$. On the other hand, we have $\kappa
Q_\MIN^2=\alpha $. Therefore if $\alpha >3(x+1)-2$, then $Q_\END <
Q_\MIN$, and we are in the desired situation. However, this is not so
simple, because the width of the hole, denoted here as $\delta (\kappa
Q^2)$, matters. We are in a good position only if $\kappa
(Q_\MIN^2-Q_\END^2) > \delta (\kappa Q^2)$; otherwise we cannot say
that the field does not feel the minimun of its potential. It is not
totally trivial to calculate the width of the potential, which is not
symmetric with respect to $Q_\MIN$. A fair estimate is given by the
difference between $Q_\MIN$ and the value of $Q$, such that the SUGRA
potential becomes different from the Ratra-Peebles (RP) potential,
i.e., for $Q$ such that $\vert V_\SUGRA (Q) / V_\RP (Q)\vert \simeq
y$, where $y$ is an arbitrary number (for example, $y = 0.1$). This
gives a width equal to $\delta (\kappa Q^2)=\alpha -2\ln (y+1)$. Of
course, the comparison depends on the precise values of $x$ and $y$,
but for reasonable values one reaches the conclusion that the width of
the potential is always of the same order of magnitude as the
difference $Q_\MIN - Q_\END$. Therefore, even if $Q_\END < Q_\MIN$, we
will obtain $\Omega_\QUINT^\OSC \ll 1$. As a consequence, the energy
density of the quintessence field cannot dominate, and there is no
possibility of reaching a value of $\Omega_\QUINT = 0.7$ today if the
minimun is put to zero.

\section{Conclusions}

In this paper, we have studied the model building problem of
quintessence in the context of SUGRA viewed as the low energy limit of
string theory. In this context, the theory is described by two scales:
the Planck scale and the string scale. {\it A priori}, there is no
reason to assume that these two scales are equal. If indeed the string
scale decouples from the Planck scale, we have shown that the SUGRA
quintessence potential arises naturally in this framework. In
addition, it was demonstrated that the potential is stable against
corrections in the K\"ahler potential and if SUSY breaking is taken
into account. A generic property of the SUGRA quintessence potentials
is the presence of a minimum. We have shown that the field today is
always close to this minimum. This requires no fine-tuning, and is due
to the fact that the minimum turns out to be of the order of the
Planck mass, the value that the field has when it leaves the
attractor, at small redshifts.  We have also demonstrated that the
minimum of the potential cannot be put to zero while keeping
$\Omega_\QUINT$ to a value of the order of the critical energy density
today.

\acknowledgements 

Section~\ref{sec_min} of this paper was motivated by numerous
interesting questions asked by Karim Benabed, Francis Bernardeau, and
Pierre Bin\'etruy. It is a pleasure to thank them for very useful
exchanges and comments. We also thank Pierre Bin\'etruy and Stephane
Lavignac for a careful reading of the manuscript. A.R. is funded by
EC-Research Training Network CMBNET (contract number
HPRT-CT-2000-00124).

\end{document}